\documentclass[cameraready]{Interspeech}

\title{Seeing the Context: Rich Visual Context-Aware Speech Recognition via Multimodal Reasoning}

\author[affiliation={1}]{Wenjie}{Tian}
\author[affiliation={1}]{Mingchen}{Shao}
\author[affiliation={1}]{Bingshen}{Mu}
\author[affiliation={1}]{Xuelong}{Geng}
\author[affiliation={1}]{Chengyou}{Wang}
\author[affiliation={1}]{Yujie}{Liao}
\author[affiliation={1}]{Zhixian}{Zhao}
\author[affiliation={1}]{Ziyu}{Zhang}
\author[affiliation={1}]{Jingbin}{Hu}
\author[affiliation={1}]{Mengqi}{Wei}
\author[affiliation={1}, correspondingauthor]{Lei}{Xie}

\address{
    $^1$ Northwestern Polytechnical University, Xi'an, China
}

\email{twj@mail.nwpu.edu.cn, lxie@nwpu.edu.cn}

\keywords{audio-visual speech recognition, multimodal large language model, multimodal reasoning, data scarcity
}

\usepackage{comment}

\begin{document}

\maketitle

\begin{abstract}
Audio-visual speech recognition (AVSR) is an extension of ASR that incorporates visual signals.
Current AVSR approaches primarily focus on lip motion, largely overlooking rich context present in the video such as speaking scene and on-screen text.
To tackle such CAVSR (AVSR including rich visual Context), we propose VASR designed to ``see" and reason the visual context to improve speech recognition.
Specifically, we construct an Audio-Visual Chain-of-Thought (AV-CoT) that explicitly enforces intermediate cross-modal grounding between acoustic signals and visual evidence.
This evidence-driven reasoning mitigates the ``single-modality dominance” problem, where models either over-rely on visual context or fail to utilize it.
Besides, to address the data scarcity, we construct and release a corresponding data pipeline and test set.
Experiments show that AV-CoT effectively mitigates the single-modality dominance, achieving state-of-the-art performance in CAVSR. 
The project is open-sourced.
\end{abstract}

\section{Introduction}
\begin{figure}[t]
    \centering
    \includegraphics[width=\linewidth]{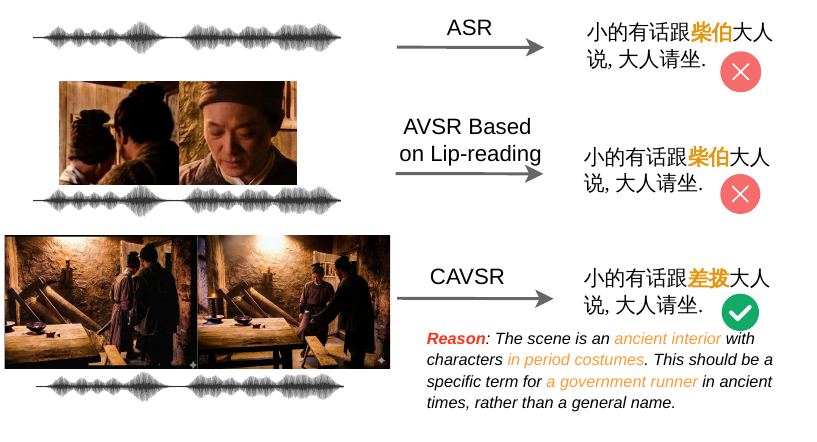}
    \caption{Comparison of different recognition tasks: ASR, AVSR focusing on lip-reading, and CAVSR focusing on rich visual context.
    This sample illustrates that the CAVSR task can leverage rich visual context to enhance recognition accuracy.}
    \label{fig:overall}
    \vspace{-0.1cm}
\end{figure}

Automatic Speech Recognition (ASR)~\cite{qwenasr, firedasr, paraformer} has achieved remarkable progress in recent years, largely driven by scaling up training data and model capacity.
However, such audio-only systems still struggle in scenarios requiring context-aware disambiguation, such as recognizing homophones, named entities, and domain-specific terms. 
This limitation largely stems from the reliance on acoustic signals alone, which provide insufficient contextual evidence.
Although some Audio-Visual Speech Recognition (AVSR) works~\cite{avsr_l_1, avsr_l_2, avsr_l_3, avsr_l_4, avsr_l_5, avsr_l_6} are proposed to include more context. 
Most of them are predominantly confined to lip movements, which require the speaker’s face to be front-facing and clearly visible.
They overlook the rich ambient visual context, such as specific scenes, objects, and captions which are universal in modern multimedia.
The rich visual context is extremely helpful to address transcription ambiguity.
For example, as shown in Figure~\ref{fig:overall}, accurately transcribing a person’s name or domain-specific terms mentioned in speech can be highly ambiguous without additional contextual cues.
This leads us to a critical question: \textbf{\textit{Can we utilize such rich visual context to improve speech recognition accuracy?}} 
We refer to this task as Context AVSR (CAVSR).

Two key aspects including model and data make the CAVSR challenging.
From the modeling perspective, recent Multimodal Large Language Models (MLLMs)~\cite{qwen2.5omni, qwen3omni, Interns1, minicpm, gemini2.5} offer a promising foundation for CAVSR because they encode the full video and audio.
However, we observe that naively applying MLLMs to CAVSR leads to systematic failures caused by ``single-modality dominance".
Specifically, when presented with on-screen subtitles that conflict with the actual acoustic evidence, the model tends to hallucinate based on the visual text, thereby disregarding the primary phonetic facts conveyed by the audio. Conversely, the MLLMs may ignore informative visual cues and rely solely on ambiguous audio.
From the data perspective, existing AVSR datasets are insufficient for studying CAVSR.
As shown in Table~\ref{tab:dataset}, most AVSR datasets primarily focus on lip motion, where the videos are often limited in face or the video’s background is frequently blurred, offering little meaningful visual context for ASR.
Other works~\cite{slideSpeech,Chineseclips}, provide video data paired with presentation slides, but these datasets are confined to highly constrained scenarios.
Existing CAVSR-related works~\cite{how2, avsr_a_1,  slideAVSR, avns, avsr_a_2} fail to address the critical challenges posed by homophone-rich languages, such as Chinese, and do not explore strategies to effectively leverage visual context for disambiguating speech.

To address these challenges in CAVSR task, we introduce \textbf{VASR} (\textbf{Visual-Aware Speech Recognition}), a framework designed to ``see" the visual context to resolve linguistic ambiguities in CAVSR.
VASR is built upon an Audio-Visual Chain-of-Thought (AV-CoT) mechanism that formulates CAVSR as a multi-step reasoning process:
\textbf{Perception} (extracting visual context and phonetic sequence), \textbf{Reasoning} (cross-modal disambiguation), and \textbf{Transcription} (generation).
By enforcing this intermediate reasoning structure, AV-CoT mitigates single-modality dominance and promotes evidence-driven fusion of audio and visual information.
Furthermore, to overcome the data scarcity in this emerging field, we developed a scalable data pipeline and curated VASR test set, a high-quality, human-verified dataset designed to test models under extreme linguistic ambiguity. We have open-sourced all of the datasets and training codes and model weights in~\url{https://github.com/wjtian-wonderful/ContextAVSR/tree/main}.
Our contributions are summarized as follows:
\begin{itemize}
\item  We propose VASR, a MLLM framework that focuses on CAVSR, shifting from local lip-reading to rich visual-aware reasoning.
\item  We propose AV-CoT, a novel multimodal reasoning process that explicitly guides MLLMs to perform cross-modal disambiguation, mitigating ``single-modality dominance".
\item  We introduce and release the VASR test set and a scalable data pipeline, providing the first comprehensive test set for evaluating CAVSR.
\item  We demonstrate through extensive experiments that VASR significantly outperforms existing strong MLLMs.
\end{itemize}

\begin{table}[!t]
  \centering
  \caption{AVSR Dataset and Corresponding Data Type}
  \label{tab:dataset}
  \begin{tabular}{lcc}  
    \toprule
    \textbf{Dataset} & \textbf{Language} & \textbf{Data Type} \\  
    \midrule
    LRS Series~\cite{lrs}       & English          & Lip-reading Videos       \\
    CMLR~\cite{cmlr}            & Chinese          & Lip-reading Videos       \\
    CN-CVS~\cite{cncvs}           & Chinese          & Lip-reading Videos       \\
    Voxceleb2~\cite{voxceleb2}        & Multilingual     & Lip-reading Videos       \\
    \midrule
    SlideSpeech~\cite{slideSpeech}      & English          & Presentation Slides\\
    Chinese-LiPS~\cite{Chineseclips}     & Chinese          & Presentation Slides\\
    \midrule
    How2~\cite{how2}             & English          & General Videos     \\
    VisSpeech~\cite{avsr_a_1}        & English          & General Videos     \\
    SlideAVSR~\cite{slideAVSR}        & English          & General Videos     \\
    AVNS~\cite{avns}             & English          & General Videos     \\
    VASR (Ours)      & Chinese          & General Videos     \\
    \bottomrule
  \end{tabular}
\end{table}

\section{Framework}

\begin{figure*}[t]
    \centering
    \includegraphics[width=\linewidth]{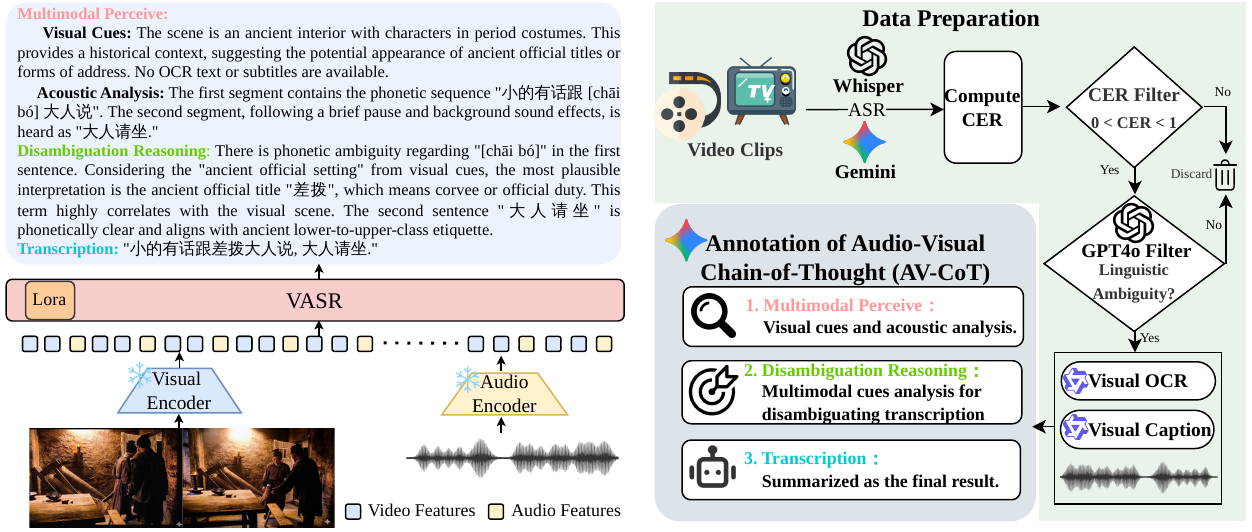}
    \caption{Overview of the VASR framework. 
    (Left) The model architecture achieves CAVSR. A qualitative example illustrates how the model resolves phonetic ambiguities via multimodal reasoning.
    (Right) The data process pipeline and construction of AV-CoT.
    }
    \label{fig:model}
\end{figure*}

\subsection{Overall Architecture}
VASR aims to achieve CAVSR in diverse real-world scenarios through structured multimodal reasoning. 
Equipped with AV-CoT, VASR reformulates CAVSR task as a structured Perception–Reasoning–Transcription pipeline. 
As illustrated in Figure~\ref{fig:model}, given multimodal inputs consisting of video $V$ and audio signals $A$, the framework first generates a reasoning path including multimodal perception, cross-modal disambiguation reasoning, and then generates transcription $\hat{Y}$.

\subsection{Multimodal Encoding}
Based on Qwen2.5-Omni's~\cite{qwen2.5omni} multimodal encoders, VASR first encodes the raw media $V$ and $A$ into continuous feature spaces, which are then interleaved temporally to form a fused multimodal stream $\mathcal{M}$. 
This unified representation $\mathcal{M}$ serves as the perceptual foundation for subsequent reasoning.

\subsection{AV-CoT mechanism}
We formulate the CAVSR task as a structured Audio-Visual Chain-of-Thought (AV-CoT) process. Given multimodal input $\mathcal{M}$, transcription is generated through a causal, autoregressive trajectory where perception and reasoning sequentially condition the final transcription generation.

\textbf{Multimodal Perception.}
AV-CoT begins with a perception stage that extracts visual context and acoustic phonetic cues.
Given multimodal input $\mathcal{M}$, the model first extracts observable visual cues like scene, on-screen text, topics, to form the visual context $C_v$.
It then extracts a phonetic sequence $P_a$ from the audio. 
For Mandarin, we use Pinyin as the basic unit.
Formally, this stage corresponds to the conditional probability of generating the perception state $S_p = \{C_v, P_a\}$ given the multimodal input $\mathcal{M}$. 
Since generation is autoregressive, it factorizes as:
\begin{equation}
P(S_p | \mathcal{M}) = \prod_{i=1}^{|C_v|} P(c_i | \mathcal{M}, c_{<i}) \prod_{j=1}^{|P_a|} P(p_j | \mathcal{M}, C_v, p_{<j}),
\end{equation}
where $c_i$ and $p_j$ denote tokens in $C_v$ and $P_a$, respectively. 
This establishes the foundation for the subsequent cross-modal disambiguation reasoning.

\textbf{Cross-modal Disambiguation.}
The reasoning stage performs cross-modal disambiguation.
Rather than directly mapping $P_a$ to text, the model generates a reasoning trajectory $R$ that aligns ambiguous phonetic spans with visual semantics $C_v$.
Formally, $R$ is generated autoregressively conditioned on $\mathcal{M}$ and the perception state:
\begin{equation}
P(R | \mathcal{M}, C_v, P_a) = \prod_{k=1}^{|R|} P(r_k | \mathcal{M}, C_v, P_a, r_{<k}).
\end{equation}
To illustrate the core disambiguation mechanism within this reasoning trajectory, consider an ambiguous phonetic span $p_{sub} \subseteq P_a$ (e.g., the syllables hé yì). 
Traditionally, audio-only systems select $y = \arg\max P(y | p_{sub})$, often defaulting to the most frequent homophone.
In contrast, AV-CoT selects from the candidate set $\mathcal{H}(p_{sub})$ by conditioning on visual context $C_v$:
\begin{equation}
y^* = \arg\max P(y \mid p_{sub}, C_v).
\end{equation}
By integrating $C_v$, the model effectively re-weights the likelihood of each candidate. 
The reasoning trajectory $R$ explicitly articulates this logical elimination by discarding grammatically incongruent options and contextually weak matches. 
This process ensures that the selected semantic representation $y^*$ is both phonetically accurate and visually grounded.


\textbf{Transcription Generation.}
The final stage, transcription, converts the inferred perception and reasoning results into transcription. 
Since visual grounding $C_v$, phonetic extraction $P_a$, and disambiguation $R$ are already modeled, decoding becomes a constrained generation process.
The transcription $\hat{Y}$ is generated autoregressively by conditioning on the entire preceding reasoning chain. 
Formally, the probability of generating the final transcription sequence of length $T$ is defined as:
\begin{equation}
P(\hat{Y} | \mathcal{M}, C_v, P_a, R) = \prod_{t=1}^{T} P(\hat{y}_t | \mathcal{M}, C_v, P_a, R, \hat{y}_{<t}),
\end{equation}
Where $\hat{y}_t$ is the $t$-th character of the final transcription. 
Here, $R$ serves as a guiding constraint on decoding.

Finally, by chaining these three stages together, the overall objective of the VASR framework can be elegantly unified. 
The joint probability of generating the perception state $S_p = \{C_v, P_a\}$, the reasoning trajectory $R$, and the final transcription $\hat{Y}$ given the multimodal input $\mathcal{M}$ is factorized according to our proposed causal dependency chain:
\begin{equation}
P(\hat{Y}, R, S_p | \mathcal{M}) = P(S_p | \mathcal{M}) \cdot P(R | \mathcal{M}, S_p) \cdot P(\hat{Y} | \mathcal{M}, S_p, R).
\end{equation}
By explicitly maximizing this joint probability during training, AV-CoT effectively forces the MLLMs to clarify what is seen and what is heard before deciding what is meant, fundamentally bridging the gap between ambient visual perception and precise linguistic decoding.

\section{Dataset }

\subsection{Data Pipeline}
To overcome the scarcity of open-source datasets for CAVSR, we design an automated, scalable data curation pipeline. 
The objective is to identify challenging, visually rich data in linguistic ambiguity and to generate high-quality multimodal annotations. 
As shown on the right of Figure~\ref{fig:model}, we split the data process into two phases: data preparation and annotation.

\textbf{Data Preparation.}
To improve data efficiency, we first employ a filtering mechanism based on acoustic ambiguity and linguistic complexity.
We perform initial ASR with two state-of-the-art (SOTA) models: Gemini2.5Pro~\cite{gemini2.5} and Whisper-Large-V3~\cite{whisper}. 
We compute the Character Error Rate (CER) between the transcriptions generated by the two models. 
To isolate challenging samples, we aggressively filter the dataset, retaining only segments where $0 < \text{CER} < 1$. 
Samples with $\text{CER} = 0$ are discarded as we assume them too simple and the audio modality alone is sufficient.
These with $\text{CER} \geq 1$ are removed to eliminate severely noisy data.
Since a non-zero CER does not guarantee the presence of linguistic ambiguity, we leverage GPT4o to analyze the transcribed text and determine its suitability for the CAVSR.

To avoid single-modality dominance during the annotation process, therefore, we explicitly extract key visual elements that matter in speech disambiguation rather than directly using raw video frames.
We use Qwen2.5-VL~\cite{qwenvl} for Optical Character Recognition (OCR). 
To differentiate between spoken text and ambient context, Qwen2.5-VL categorizes the on-screen text into two distinct classes: spoken subtitles and background text such as watermarks, brand logos, or text printed on objects.
Meanwhile, we employ Qwen2.5-VL to generate a comprehensive video caption of the visual scene, capturing the environment, objects, and actions present in the video.

\textbf{AV-CoT Annotation.}
After obtaining the video annotations, we feed the raw audio and visual annotations into Gemini2.5Pro to generate the reasoning path.
We first perform multimodal perception.
For the visual modality, we summarize OCR and video captions into visual cues.
For acoustic analysis, we extract phonetic sequences and identify ambiguous segments using the pronunciation sequence (i.e., Pinyin in Chinese).
Providing more linguistic and situational context allows the model to infer the most likely words in unclear or ambiguously pronounced speech segments.
We then leverage the generated visual cues to disambiguate the ambiguous segments in the pronunciation sequence, thereby yielding the final transcription.


\begin{table*}[!ht]
  \centering
  \caption{Comparison of model performance. 
  The best and second-best results are shown in bold and underlined, respectively.
  }
  \label{tab:table_exp}
  \begin{tabular}{lccccc}  
    \toprule
    Model & Parameters & \shortstack{Input Modality} & \shortstack{Task} & \multicolumn{2}{c}{CER(\%) $\downarrow$} \\ 
    \cmidrule(lr){5-6} 
          &            &                            &                  & Chinese-LiPS & VASR Test \\ 
    \midrule
    Gemini2.5Pro~\cite{gemini2.5}                    & -     & A/V         & CAVSR                      & 4.59  & \underline{11.81} \\
    Intern-S1~\cite{Interns1}                        & 9B    & A/V         & CAVSR                      & 71.94    & 17.6 \\
    MiniCPM-o2.6~\cite{minicpm}                      & 8B    & A/V         & CAVSR                      & 73.92 & 19.49 \\
    Qwen3Omni-Instruct-30B-A3B~\cite{qwen3omni}      & 30B   & A/V         & CAVSR                      & \underline{4.41}  & 11.97 \\
    Qwen3Omni-Thinking-30B-A3B~\cite{qwen3omni}      & 30B   & A/V         & CAVSR                      & 7.56    & 12.39 \\
    Qwen2.5Omni-7B~\cite{qwen2.5omni}                & 7B    & A/V         & CAVSR                      & 22.45    & 12.21 \\
    Qwen2.5Omni-3B~\cite{qwen2.5omni}                & 3B    & A/V         & CAVSR                      & 27.89 & 13.06 \\
    \midrule
    Doubao ASR                                       & -     & A           & ASR                       & 3.47  & 17.17 \\
    Qwen2.5Omni-7B                                       & 7B     & A           & ASR                       & 11.71  & 19.25 \\
    \midrule

    VASR                                           & 7B    & A/V         & CAVSR                      & \textbf{1.80}  & \textbf{11.02} \\
    \bottomrule
  \end{tabular}
\end{table*}

\begin{table}[!ht]
\centering
\caption{
Ablation study to validate the contribution of the AV-CoT. 
``w/o AV-CoT": direct SFT without AV-CoT; ``w/ Black Video": black (all-zero frames) video used in inference; 
``w/ Random Video": random video used in inference.
}
\label{tab:abl}
\begin{tabular}{lcc} 
\toprule
Model/Metrics & \multicolumn{2}{c}{CER(\%) $\downarrow$} \\ 
\cmidrule(lr){2-3} 
              & Chinese-LiPS & VASR Test \\ 
\midrule
VASR & 1.80  & 11.02 \\
\quad w/o AV-COT          & 2.65  & 12.66 \\ 
\quad w/ Black Video          & 4.19  & 16.25 \\ 
\quad w/ Random Video          & 5.29  & 18.94 \\ 

\bottomrule
\end{tabular}
\vspace{-0.2cm}
\end{table}

\subsection{VASR Test Set}
We randomly sample 2,000 utterances from the used datasets, manually verify the corresponding transcripts, and construct the final CAVSR test set.
Meanwhile, we manually check whether each sample is suitable for the CAVSR task and remove overly simple samples. During manual filtering, the reasons generated by GPT are provided for human reference.
The final test set consists of 1,981 utterances.

\section{Experiments}

\subsection{Datasets}
We utilize open-source chinese datasets, including MEIJU~\cite{MEIJU} and MER25~\cite{mer25}, ch-sims-v2~\cite{chsimsv2},
and the Chinese-LiPS~\cite{Chineseclips} training set. 
All open-source data are uniformly processed using our proposed data processing pipeline, totaling approximately 230 hours.
We use Chinese-LiPS test set as the open-source test set. For general-scenario, we also include our proposed VASR test set as a more comprehensive and challenging dataset.


\subsection{Implementation Details}
We adopt Qwen2.5-Omni-7B as the base model for all experiments. 
The visual encoder and audio encoder of VASR are frozen throughout the training phase.
All training experiments are executed on 16 NVIDIA A800 GPUs with mixed-precision enabled.
We finetune the thinker module using LoRA for three epochs, with a total batch size of 96 and a learning rate of $1 \times 10^{-4}$. The LoRA configuration employs a rank of 8, an alpha value of 32, and a dropout rate of 0.05.
In all experiments, we employ a cosine learning rate scheduler with 3,000 warm-up steps. 
All inputs are preprocessed using the Qwen2.5-Omni's processor to ensure compatibility with the base model.

\subsection{Comparison Models and Evaluation Metric.}
Since several existing unconstrained AVSR models are not open-sourced, we can only compare our baseline Qwen2.5-Omni-7B~\cite{qwen2.5omni} and several strong Omni models: Qwen3-Omni-30B-A3B-Instruct~\cite{qwen3omni}, Qwen3-Omni-30B-A3B-Thinking~\cite{qwen3omni}, Intern-S1-9B~\cite{Interns1}, and MiniCPM-o2.6-8B~\cite{minicpm}.
We also include two strong commercial baselines: Gemini2.5Pro~\cite{gemini2.5} and Doubao ASR~\footnote{\url{https://www.volcengine.com/docs/6561/1354869?lang=en}}.
We adopt CER to measure the character error rate in Chinese.

\subsection{Experiments Results}
Table~\ref{tab:table_exp} presents the performance comparison of VASR against SOTA commercial and open-source models, as well as a robust audio-only baseline. 
Our proposed VASR consistently achieves SOTA results across diverse scenarios, ranging from dense-text videos to visually rich videos. 
It outperforms all baselines by a significant margin, including large-scale models such as Gemini2.5Pro and Qwen3Omni series. 
Remarkably, VASR is built on a Qwen2.5omni-7B backbone and fine-tuned using LoRA with only a few hundred hours of data. 
This underscores the simplicity and remarkable effectiveness of the AV-CoT method in tackling the CAVSR task.
Although the ASR performance of Qwen2.5Omni-7B is inferior to that of Doubao ASR, our proposed VASR model effectively leverages visual context and further elevates the upper bound of the recognition task.
By comparing the ASR and CVASR results of Qwen2.5Omni-7B, we observe that without addressing the issue of single-modality dominance, rich visual context tends to degrade rather than improve the model’s recognition performance.

A striking anomaly is observed with Intern-S1 and MiniCPM-o2.6 on the Chinese-LiPS dataset, where they yield extremely high CERs of 71.94\% and 73.92\%, respectively. We attribute this to the fact that they are heavily distracted by the dense text on the slides.
This failure mode underscores a critical limitation in addressing single-modality dominance. 
\subsection{Ablation Studies}
To validate the contribution of AV-CoT, we finetune the model on data without AV-CoT.
As shown in Table~\ref{tab:abl}, this leads to an increase in CER across both benchmarks, indicating that the model fails to fully exploit visual context.

To further verify that AV-CoT indeed leverages visual context to enhance recognition accuracy,
we keep the inference audio unchanged:
one variant uses a randomly selected video (w/ random video),
and the other uses a video with all-zero frames (w/ black video).
The results with black video show that removing visual context leads to a performance degradation, indicating that the model actively leverages visual cues rather than ignoring the visual modality.
Meanwhile, the two variants' performance drops to a level comparable to Doubao ASR as expected, while still outperforming most multimodal baselines. 
This indicates that the model does not over-rely on potentially misleading visual signals and maintains stable performance even when visual information is corrupted, suggesting that AV-CoT helps alleviate the single-modality dominance issue.
\section{Conclusion and Limitation}
In this work, we introduce the CAVSR task, extending AVSR beyond lip-reading to leverage rich visual context for resolving linguistic ambiguities. 
We propose VASR, which leverages AV-CoT to explicitly model the transcription process to address the single-modality dominance. 
In addition, we develop and release a scalable data construction pipeline and introduce the VASR test set to facilitate a systematic study of the CAVSR task. 
Extensive experiments demonstrate that VASR achieves state-of-the-art performance and significantly outperforms other strong MLLMs.
However, the visual encoder's frame rate in the pretrained Qwen2.5-Omni model is so low that we cannot integrate the lip-reading task into our work.

\newpage
\section{Generative AI Use Disclosure}
Generative AI tools are used for writing refinement and data generation. They play no role in methodology, experimentation, interpretation, or the production of scientific results. The authors bear full intellectual responsibility for all content in this manuscript.
All authors agree with the submission of this paper.

\bibliographystyle{IEEEtran}
\bibliography{mybib}

\end{document}